\documentclass{article}                  %% LaTeX 2e
\usepackage{opex3}
\usepackage{graphicx}

\begin{document}

\title{Linear and nonlinear dynamics of matter wave packets in periodic potentials}

\author{Th. Anker, M. Albiez, B. Eiermann, M. Taglieber and M.~K. Oberthaler}
\address{Kirchhoff Institut f\"ur Physik, Universit\"at Heidelberg, Im Neuenheimer Feld 227, 69120 Heidelberg}
\email{thomas.anker@kip.uni-heidelberg.de} \homepage{http://www.kip.uni-heidelberg.de/matterwaveoptics}

\begin{abstract}
We investigate experimentally and theoretically the nonlinear
propagation of  $^{87}$Rb Bose Einstein condensates in a trap with
cylindrical symmetry. An additional weak periodic potential which
encloses an angle with the symmetry axis of the waveguide is
applied. The observed complex wave packet dynamics results from
the coupling of transverse and longitudinal motion. We show that
the experimental observations can be understood applying the
concept of effective mass, which also allows to model numerically
the three dimensional problem with a one dimensional equation.
Within this framework the observed slowly spreading wave packets
are a consequence of the continuous change of dispersion. The
observed splitting of wave packets is very well described by the
developed model and results from the nonlinear effect of transient
solitonic propagation.
\end{abstract}
\vspace{5mm}
%\ocis{(270.5530) Pulse propagation and solitons; (020.0020) Atomic
%and molecular physics; (350.4990) Particles}

% NOTE: \ocis{} COULD/SHOULD BE ALIASED TO \pacs{} BUT
                          % MUST FORMAT THE TERMS CORRECTLY FOR OPEX

%\maketitle               % \maketitle IS A NULL FUNCTION in OPEX

\section{Introduction}

The experimental investigation of nonlinear matter wave dynamics
is feasible since the realization of Bose-Einstein-condensation of
dilute gases \cite{BEC_general}. The combination of this new
matter wave source with periodic potentials allows for the
realization of many nonlinear propagation phenomena. The dynamics
depends critically on the modulation depth of the potential. For
deep periodic potentials the physics is described locally taking
into account mean field effects and tunneling between adjacent
potential wells. In this context wave packet dynamics in Josephson
junction arrays  have been demonstrated experimentally
\cite{Inguscio} and nonlinear self trapping has been predicted
theoretically \cite{Trombettoni01}. In the limit of weak periodic
potentials and moderate nonlinearity rich wave packet dynamics
result due to the modification of dispersion which can be
described applying band structure theory \cite{Steel98}.
Especially matter wave packets subjected to anomalous dispersion
(negative effective mass) or vanishing dispersion (diverging mass)
are of great interest. In the negative mass regime gap solitons
have been predicted theoretically \cite{gapsoliton} and have been
observed recently \cite{obergapoliton}. Also modulation
instabilities can occur \cite{modulationinstability}.

The experiments described in this work reveal wave dynamics in the
linear and nonlinear regime for weak periodic potentials. The
observed behavior is a consequence of the special preparation of
the wave packet leading to a continuous change of the effective
mass and thus the dispersion during the propagation. The initial
propagation is dominated by the atom-atom interaction leading to
complex wave dynamics. After a certain time of propagation slowly
spreading atomic wave packets are formed which are well described
by linear theory. In this work we focus on the mechanisms
governing the initial stage of propagation.

The paper is organized as follows: in section
{\ref{Effectivemassconcept}} we describe the effective mass and
dispersion concept. In section {\ref{ExperimentalSetup}} we
present our experimental setup and in section {\ref{preparation}}
the employed wave packet preparation schemes are discussed in
detail. In section {\ref{ExperimentalResults}} the experimental
results are compared with numerical simulations. We show that some
features of the complex dynamics can be identified with well known
nonlinear mechanisms. We conclude in section {\ref{Conclusion}}.

\section{Effective mass and dispersion concept}\label{Effectivemassconcept}

\begin{figure}[h!]
\centerline{\includegraphics[width=8cm]{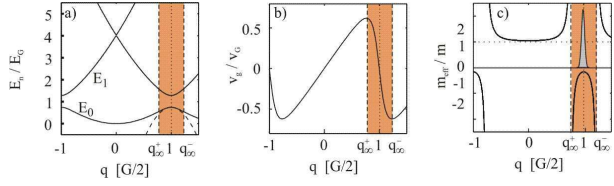}}
\caption{\label{fig:1}(a) Band structure for atoms in an optical
lattice with $V_0=1.2\,E_{rec}$ (solid), parabolic approximation
to the lowest energy band at $q=\pi/d=G/2$ (dashed), corresponding
group velocity (b) and effective mass (c) in the lowest energy
band. The vertical dashed lines at $q=q^\pm_\infty$ indicate where
$|m_\mathrm{eff}|=\infty$. The shaded region shows the range of
quasimomenta where the effective mass is negative.}
\end{figure}

In our experiments we employ a weak periodic potential which leads
to a dispersion relation $E_n(q)$ shown in Fig.~\ref{fig:1}(a).
This relation is well known in the context of electrons in
crystals \cite{Ashcroftenglish76} and exhibits a band structure.
It shows the eigenenergies of the Bloch states as a function of
the quasi-momentum $q$. The modified dispersion relation leads to
a change of wavepacket dynamics due to the change in group
velocity $v_g(q)=1/\hbar \;\partial E/\partial q$ (see Fig.
\ref{fig:1}(b)), and the group velocity dispersion described by
the effective mass $m_{eff}=\hbar^2(\partial^2 E/\partial
q^2)^{-1}$ (see Fig. \ref{fig:1}(c)), which is equivalent to the
effective diffraction introduced in the context of light beam
propagation in optically-induced photonic lattices
\cite{Sukhorukov}. In our experiment only the lowest band is
populated, which is characterized by two dispersion regimes,
normal and anomalous dispersion, corresponding to positive and
negative effective mass. A pathological situation arises at the
quasimomentum $q_\infty^{\pm}$, where the group velocity $v_g(q)$
is extremal, $|m_{eff}|$ diverges and thus the dispersion
vanishes.

In the following we will show that the two preparation schemes
employed in the experiment lead to a continuous change of the
quasimomentum distribution, and thus to a continuous change of
dispersion. One of the preparation schemes allows to switch
periodically from positive to negative mass values and thus a
slowly spreading wave packet is formed. This is an extension of
the experiment reporting on dispersion management
\cite{Eiermann1}. The second preparation gives further insight
into the ongoing nonlinear dynamics for the initial propagation.

\section{Experimental Setup}\label{ExperimentalSetup}

\begin{figure}[h!]
\centerline{ \includegraphics[width=8cm]{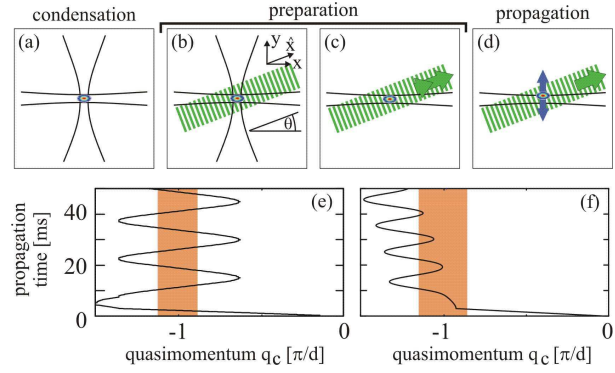}}
\caption{\label{fig:2} Scheme for wave packet preparation (a-d).
(a) initial wave packet is obtained by condensation in a crossed
dipole trap. (b) A stationary periodic potential is ramped up
adiabatically preparing the atoms at quasimomentum $q_c=0$ in the
lowest band. (c),(d) The periodic potential is accelerated to a
constant velocity. (e) shows the numerically deduced quasimomentum
shift for the preparation method I described in the text. (f) The
motion of the center quasimomentum for the preparation method II
described in the text. The additional shift to higher quasimomenta
for long times results from the residual trap in the direction of
the waveguide. The shaded area represents the quasimomenta
corresponding to negative effective mass.}
\end{figure}

The wave packets in our experiments have been realized with a
$^{87}$Rb Bose-Einstein condensate (BEC). The atoms are collected
in a magneto-optical trap and subsequently loaded into a magnetic
time-orbiting potential trap. By evaporative cooling we produce a
cold atomic cloud which is then transferred into an optical dipole
trap realized by two focused Nd:YAG laser beams with $60\,\mu m $
waist crossing at the center of the magnetic trap (see
Fig.\ref{fig:2}(a)). Further evaporative cooling is achieved by
lowering the optical potential leading to pure Bose-Einstein
condensates with $1 \cdot 10^4$ atoms in the $|F=2, m_F=+2\rangle$
state. By switching off one dipole trap beam the atomic matter
wave is released into a trap acting as a one-dimensional waveguide
with radial trapping frequency $\omega_\perp = 2 \pi \cdot 100
\,Hz$ and longitudinal trapping frequency $\omega_\parallel = 2
\pi \cdot 1.5 \,Hz$. It is important to note that the dipole trap
allows to release the BEC in a very controlled way leading to an
initial mean velocity uncertainty smaller than 1/10 of the photon
recoil velocity.

The periodic potential is realized by a far off-resonant standing
light wave with a single beam peak intensity of up to $1\,W/cm^2$.
The chosen detuning of 2\,nm to the blue off the D2 line leads to
a spontaneous emission rate below $1\, Hz$. The standing light
wave and the waveguide enclose an angle of $\theta=21^\circ$ (see
Fig.~\ref{fig:2}(b)). The frequency and phase of the individual
laser beams are controlled by acousto-optic modulators driven by a
two channel arbitrary waveform generator allowing for full control
of the velocity and amplitude of the periodic potential. The light
intensity and thus the absolute value of the potential depth was
calibrated independently by analyzing results on Bragg scattering
\cite{Bragg} and Landau Zener tunneling
\cite{Kasevich,Morsch01b,Raizen97}.

The wave packet evolution inside the combined potential of the waveguide and the lattice is studied by taking
absorption images of the atomic density distribution after a variable time delay. The density profiles along the
waveguide, $n(x,t)$, are obtained by integrating the absorption images over the transverse dimension.

\section{Dynamics in reciprocal space}\label{preparation}

In our experimental situation an acceleration of the periodic
potential to a constant velocity leads to a collective transverse
excitation as indicated in Fig.~\ref{fig:2}(d). Since the
transverse motion in the waveguide has a non vanishing component
in the direction of the periodic potential due to the angle
$\theta$, a change of the transverse velocity leads to a shift of
the central quasimomentum of the wave packet. The coupling between
the transverse motion in the waveguide and the motion along the
standing light wave gives rise to a nontrivial motion in
reciprocal (see Fig.~\ref{fig:2}(e,f)) and real space.

The appropriate theoretical description of the presented
experimental situation requires the solution of the three
dimensional nonlinear Schr\"odinger equation (NLSE) and thus
requires long computation times. In order to understand the basic
physics we follow a simple approach which solves the problem
approximately and explains all the features observed in the
experiment. For that purpose we first solve the semiclassical
equations of motion of a particle which obeys the equation
$\vec{F} = M^* \ddot{\vec{x}} $ where $M^*$ is a mass tensor
describing the directionality of the effective mass. We deduce the
time dependent quasimomentum $q_c(t)$ in the direction of the
periodic potential by identifying $\hbar \dot{q_c} = F_{\hat x}$
and $\dot{\hat{x}} = v_g(q_c)$ (definition of $\hat{x}$ see
Fig.~\ref{fig:2}(b)). Subsequently we can solve the one
dimensional NPSE (non-polynomial nonlinear Schr\"odinger
equation)\cite{Salasnich} where the momentum distribution is
shifted in each integration step according to the calculated
$q_c(t)$. Thus the transverse motion is taken into account
properly for {\em narrow} momentum distributions. We use a split
step Fourier method to integrate the NPSE where the kinetic energy
contribution is described by the numerically obtained energy
dispersion relation of the lowest band $E_0(q)$. It is important
to note, that this description includes all higher derivatives of
$E_0(q)$, and thus goes beyond the effective mass approximation.

In the following we discuss in detail the employed preparation schemes:

{\em Acceleration scheme I}: After the periodic potential is
adiabatically ramped up to $V_0=6 E_{rec}$ it is accelerated
within $3\,$ms to a velocity $v_{pot}=\cos^2(\theta)1.5v_{rec}$.
Then the potential depth is lowered to $V_0=0.52 E_{rec}$ within
$1.5\,$ms and the periodic potential is decelerated within $3\,$ms
to $v_{pot}=\cos^2(\theta)v_{rec}$ subsequently. $V_0$ and
$v_{pot}$ are kept constant during the following propagation. The
calculated motion in reciprocal space $q_c(t)$ is shown in
Fig.~\ref{fig:2}(e).

{\em Acceleration scheme II}: The periodic potential is ramped up
adiabatically to $V_0=0.37\,E_{rec}$ and is subsequently
accelerated within $3\,$ms to a final velocity
$v_{pot}=\cos^2(\theta) \times 1.05\, v_{rec}$. The potential
depth is kept constant throughout the whole experiment.
Fig.~\ref{fig:2}(f) reveals that in contrast to the former
acceleration scheme the quasimomentum for the initial propagation
is mainly in the negative effective mass regime.

\section{Experimental and Numerical Results}\label{ExperimentalResults}

In this section we compare the experimental results with the
predictions of our simple theoretical model discussed above. The
numerical simulation reveal all the experimentally observed
features of the dynamics such as linear slowly spreading
oscillating wave packets, nonlinear wave packet compression and
splitting of wave packets. The observed nonlinear phenomena can be
understood by realizing that in the negative effective mass regime
the repulsive atom-atom interaction leads to compression of the
wave packet in real space and to a broadening of the momentum
distribution. An equivalent picture borrowed from nonlinear photon
optics \cite{Agrawal01} is the transient formation of higher order
solitons, which show periodic compression in real space with an
increase in momentum width and vice versa.

\subsection{Preparation I}
\begin{figure}[h!]
\centerline{ \includegraphics[width=8cm]{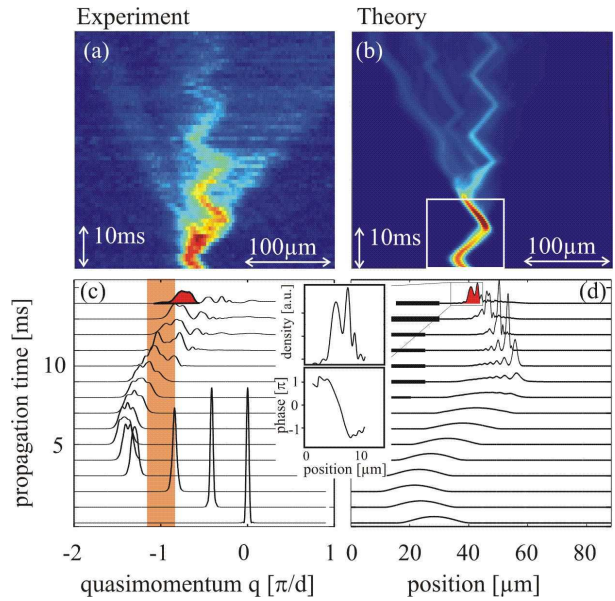}}
\caption{\label{fig:3}Wave packet dynamics for preparation I. (a)
Experimental observation of wave packet propagation. (b) Result of
the numerical simulation as discussed in the text. The data is
convoluted with the optical resolution of the experiment. The
obtained results are in good agreement with the experimental
observations. The theoretically obtained (c) quasimomentum
distribution and (d) real space distribution are given for the
initial 14ms of propagation. The inset reveals the phase of the
observed slowly spreading wave packet.}
\end{figure}

The experimental results for the first acceleration scheme
discussed in section {\ref{preparation}} are shown in Fig.
\ref{fig:3}. Clearly we observe that a wave packet with reduced
density is formed which spreads out slowly and reveals
oscillations in real space. This wave packet results from the
initial dynamics characterized by two stages of compression which
lead to radiation of atoms \cite{scott}. The observed behavior is
well described by our numerical simulation which allows further
insight into the ongoing physics.

In Fig. \ref{fig:3}(c,d) we show the calculated momentum and real
space distribution for the first 14ms of propagation. As can be
seen the acceleration of the standing light wave leads to a
oscillatory behavior in momentum space. For the chosen parameters
the wave packet is initially dragged with a tight binding
potential ($V_0=6 E_{rec}$) over the critical negative mass
regime. While the real space distribution does not change during
this process, the momentum distribution broadens due to self phase
modulation \cite{Agrawal01}. The subsequent propagation in the
positive mass regime leads to a further broadening in momentum
space and real space (t=4-9ms).

The dynamics changes drastically as soon as a significant part of
the momentum distribution populates quasimomenta in the negative
mass regime (t=10ms). There the real space distribution reveals
nonlinear compression as known from the initial dynamics of higher
order solitons. This compression leads to a significant further
broadening in momentum space and thus to population of
quasimomenta corresponding to positive mass. This results in a
spreading in real space due to the different group velocities
involved and leads to the observed background. The change of the
quasimomentum due to the transverse motion prohibits a further
significant increase in momentum width, since the whole momentum
distribution is shifted out of the critical negative mass regime
at t=14ms.

The long time dynamics of the slowly spreading wave packet is
mainly given by the momentum distribution marked with the shaded
area for t=14ms in Fig. \ref{fig:3}(c). The subsequent motion is
dominated by the change of the quasimomentum due to the transverse
motion. This leads to a periodic change from normal to anomalous
dispersion and thus the linear spreading is suppressed. This is an
extension of our previous work on dispersion management for matter
waves - continuous dispersion management.

\subsection{Preparation II}
\begin{figure}[h!]
\centerline{ \includegraphics[width=8cm]{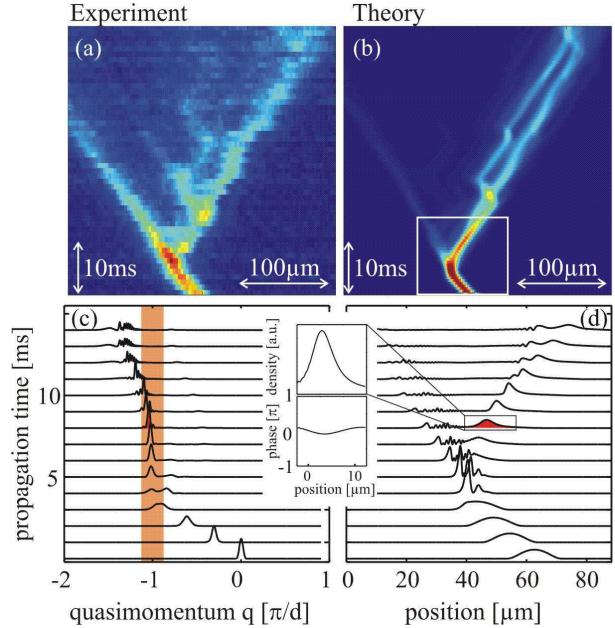}}
\caption{\label{fig:4}Wave packet dynamics for preparation II. (a)
Experimental results on wave packet propagation. (b) Result of the
numerical simulation as discussed in the text. The simulation
reproduces the observed wave packet splitting. The theoretically
obtained (c) quasimomentum distribution and (d) real space
distribution are given for the initial 14ms of propagation. The
inset reveals that the transient formed wave packet has a flat
phase indicating solitonic propagation.}
\end{figure}

This preparation scheme reveals in more detail the transient
solitonic propagation leading to the significant spreading in
momentum space. This results in a splitting of the wave packet
which cannot be understood within a linear theory. The results are
shown in Fig. \ref{fig:4} and the observed splitting is confirmed
by our numerical simulations.

In contrast to the former preparation scheme the momentum
distribution is prepared as a whole in the critical negative mass
regime. Our numerical simulations reveal that the wave packet
compresses quickly in real space after t=4ms which is accompanied
by an expansion in momentum space. The momentum distribution which
stays localized in the negative mass regime reveals further
solitonic propagation characterized by an expansion in real space
and narrowing of the momentum distribution (t=5-10ms). The
transverse motion shifts this momentum distribution into the
normal dispersion regime after 11ms of propagation resulting in a
wave packet moving with positive group velocity (i.e. moving to
the right in fig. {\ref{fig:4}}(b)). The initial compression at
t=4ms even produces a significant population of atoms in the
normal mass regime which subsequently move with negative group
velocity showing up as a wave packet moving to the left in Fig.
{\ref{fig:4}}(b). Thus the splitting in real space is a
consequence of the significant nonlinear broadening in momentum
space.

\section{Conclusion}\label{Conclusion}
In this paper we report on experimental observations of nonlinear
wave packet dynamics in the regime of positive and negative
effective mass. Our experimental setup realizing a BEC in a
quasi-one dimensional situation allows the observation of wave
dynamics for short times, where the nonlinearity due to the
atom-atom interaction dominates and also for long times, where
linear wave propagation is revealed.

We have shown that a slowly spreading wave packet can be realized
by changing the quasimomentum periodically from the normal to
anomalous dispersion regime. This can be viewed as an
implementation of continuous dispersion management. We further
investigate in detail the formation process of these packets,
which are a result of the initial spreading in momentum space due
to nonlinear compression. A second experiment investigates in more
detail the nonlinear dynamics in the negative mass regime  where
the solitonic propagation leads to a significant broadening in
momentum space. This shows up in the experiment as splitting of
the condensate into two wave packets which propagate in opposite
directions.

The developed theoretical description utilizing the effective mass tensor models the experimental system in one
dimension and can explain all main features observed in the experiment.

\section{Acknowledgment}
This work was supported by Deutsche Forschungsgemeinschaft, Emmy Noether Program, by the European Union, Contract No.
HPRN-CT-2000-00125, and the Optik Zentrum University of Konstanz.


\begin{thebibliography}{}
\bibitem{obergapoliton}
The experimental realization in our group will be published
elsewhere.
\end{thebibliography}
\end{document}